\documentclass[aps, prb, 10pt, twocolumn, superscriptaddress, longbibliography, nofootinbib]{revtex4-2}
\usepackage[utf8]{inputenc}
\usepackage[english]{babel}
\usepackage[caption=false]{subfig}
\usepackage{amsmath,graphicx,bm}
\usepackage{xcolor}
\usepackage{booktabs}
\usepackage{multirow}
\usepackage{microtype}
\usepackage[plainpages=false,colorlinks=true,linkcolor=blue, citecolor=blue, urlcolor=blue]{hyperref}
\usepackage{cleveref}

\renewcommand\Im{\,\mathrm{Im}}

\newcommand\dg{\dagger}

\newcommand\bc{\mathbf{c}}
\renewcommand\bf{\mathbf{f}}
\newcommand\bH{\mathbf{H}}

\begin{document}
\title{Computing Green's functions and improving ground state energy estimation on quantum computers with Liouvillian recursion}

\author{J. Leblanc}
\author{O. Nahman-L\'evesque}
\author{J. Forget}
\author{T. Lepage-L\'evesque}
\author{S. Verret}
\email[Corresponding author: ]{simon.verret@usherbrooke.ca}
\author{A. Foley}
\affiliation{Algolab, Institut quantique, Universit\'e de Sherbrooke, Sherbrooke, QC, Canada}
\date{\today}

\begin{abstract}
We present a quantum-classical hybrid implementation of the Liouvillian recursion method to compute many-body Green's functions using a quantum computer. 
From an approximate ground state preparation circuit, this algorithm produces the local ($r=r'$) and inter-site ($r\neq r'$) Green's functions $G_{rr'}(\omega)$ by measuring observables generated recursively.
We demonstrate the approach on a superconducting quantum processor for the open-boundary four-site Hubbard model.
We then use the computed Green's functions as input to the Galitskii-Migdal formula to produce better ground state energy estimation than the expectation value of the Hamiltonian for the approximate circuit.
Empirical results indicate exponential convergence in the number of iterations, yielding a computational complexity polynomial in the Green's-function accuracy, as measured with the Wasserstein distance.
Our results also indicate significant robustness to noise and to inaccuracies of the ground state preparation, providing evidence that Liouvillian recursion is well adapted to the constraints of near-term quantum computing.
\end{abstract}

\maketitle

One of the most promising applications of quantum computing is the simulation of strongly correlated electron systems~\cite{lloyd_Universal_1996,somma_simulating_2002,aspuru-guzik_Simulated_2005,bauer_Hybrid_2016}. Candidates for early quantum advantage include small molecules and cluster impurity models, which underpin dynamical mean-field theory (DMFT) approaches to correlated materials. While determining the ground state of such systems is of intrinsic interest, extracting physical observables and leveraging DMFT typically requires response functions like the single-particle Green’s function.

Many quantum algorithms have been proposed to compute the Green's functions.
To date, all such algorithms require some form of ground state preparation, and most of them rely on additional costly operations, such as multi-qubits control operations~\cite{dallaire-demers_Method_2016,dallaire-demers_Quantum_2016,baker_Lanczos_2021,tong_Fast_2021,bishop_Quantum_2025} or time evolution~\cite{jamet_Quantum_2022,gomes_computing_2023,li_utilizing_2024,piccinelli_Efficient_2025,singh_Classical_2026}. This leads to quantum circuits of depth prohibitive for near-term quantum computers.
A few algorithms, however, only require query access to the ground state, circumventing the need for additional operations~\cite{gauthier_Occupationnumber_2024, selisko_Dynamical_2024,greene-diniz_Quantum_2024,irmejs_Approximating_2025}.
They can be divided in two different approaches.
The first projects the Hamiltonian in a subspace and builds the Green's function in the Lehmann representation through subspace diagonalization.
Examples include the works of \citet{gauthier_Occupationnumber_2024}, who use non-orthogonal basis and highlight opportunities for noise filtering in building a low-excitation subspace, and  \citet{selisko_Dynamical_2024}, who showcase how to use such method as an impurity solver in cluster extensions to DMFT.
The second approach uses recursion relations or moment expansions to build the Green's function in a continued-fraction representation. 
These approaches are related to the Lanczos method~\cite{dagotto_Correlated_1994}, and more systematic than other subspace methods. They fit the broader framework of the recursion method~\cite{viswanath_Recursion_1994}, with various formulations in the Hamiltonian formalism and the Liouvillian formalism.
Examples include the works of \citet{greene-diniz_Quantum_2024}, who showcase a DMFT impurity solver where the Green's function is built by measuring moments of the Hamiltonian, and \citet{irmejs_Approximating_2025}, who use the projection operator formalism to compute Green's functions and prove the exponential convergence of the Liouvillian recursion with respect to the number of iterations. 
They also prove stability to noise and ground state imperfections; a particularly important feature when considering that ground state preparation can be hard even for a quantum computer~\cite{kempe_Complexity_2006}.

In this paper, we present a new quantum algorithm based on the Liouvillian recursion approach, and
we show that, given query access to an \emph{approximate} ground state, we can compute accurate Green's functions and improve this approximate ground state's energy estimation.
The paper is divided as follows.
First, we explain how to obtain a continued fraction representation for the one-particle Green's functions in the frequency domain by measuring a sequence of observables generated by Liouvillian recursion.
Next, we demonstrate the viability of the approach on the {\it IBM Quebec} superconducting quantum processor, obtaining accurate Green's functions for the four-site open boundary Fermi-Hubbard model.
We then show how to estimate the true ground state energy from the calculated Green's functions and we find that this estimate is more accurate than the energy expectation value of the approximately prepared ground state.
Finally, we investigate the computational cost of our algorithm. We show that although the number of Pauli operators generated by the recursion grows exponentially with the number of iterations, the exponential convergence of the recursion method compensates this cost. This leads to an effective complexity that is polynomial in the Green's function accuracy, where the accuracy is measured via the Wasserstein distance.
Throughout this work, we highlight evidence of the algorithm's robustness to noise and to inaccuracies in the ground state, suggesting a natural affinity between the Liouvilian recursion method and near-term quantum computation.

Our algorithm is similar in spirit to that of~\citet{irmejs_Approximating_2025}, but differs in two key aspects. Firstly, we obtain the full retarded Green's function, rather than separate hole and electron contributions. Secondly, we obtain the off-diagonal elements using a polynomial hybrid formulation instead of with a continued fraction of matrices.
Our approach was detailed by~\citet{foley_liouvillian_2024} for the classical case; here we present an implementation adapted to quantum computers.

To begin, we consider the retarded Green's function,
\begin{equation}
    G_{rr'}(t) = -i \langle \{ \bc_{r}^{\phantom{\dagger}}(t), \bc_{r'}^\dagger \}  \rangle \theta(t),
\label{eq:GreenT}
\end{equation}
where $\bc_{r}(t)$ and $\bc_{r}^{\dg}(t)$ are respectively the electronic annihilation and creation operators at time $t$ and state $r$, a composite index that includes site $i$ and spin $\sigma$, and $\theta(t)$ is the Heaviside step function. The average $\langle \cdots \rangle$ is with respect to the zero temperature ground state.
We obtain the diagonal Green's functions ($r'=r$) as a continued fraction representation in the frequency~$\omega$ domain,
\begin{equation}
    G^{(k)}_{rr}(\omega) = \cfrac{1}{\omega - \alpha_{0,r} 
- \cfrac{\beta_{1,r}^2}{\omega- \alpha_{1,r} -
\raisebox{-0.6em}{$
\ddots
\raisebox{-0.3em}{$-\dfrac{\beta_{k,r}^2}{\omega- \alpha_{k,r} }$}
$}
}}
,
\label{eq:contfrac}
\end{equation}
where the coefficients $\alpha_{k,r}$ and $\beta_{k,r}$ are computed recursively up to iteration~$k$, using the following set of equations:
\begin{align}
\beta_{k+1,r}\bf_{k+1,r} &= [\bH,\bf_{k,r}] - \alpha_{k,r} \bf_{k,r} - \beta_{k,r}\bf_{k-1,r}, 
\label{eq:recurs}\\
\beta^2_{k,r} &= \langle \{\beta_{k,r}\bf_{k,r}^\dagger, \beta_{k,r}\bf_{k,r} \}\rangle,
\label{eq:a}\\
\alpha_{k,r} &= \langle \{\bf_{k,r}^\dagger , [\bH,\bf_{k,r}]\} \rangle,
\label{eq:b}
\\
\bf_{0,r} &= \bc_{r},
\label{eq:f0}\\
\bf_{-1,r} &= 0.
\label{eq:fm1}
\end{align}
This algorithm is equivalent to the Lanczos algorithm~\cite{dagotto_Correlated_1994}, but as applied for a Hilbert space of \emph{operators} instead of a Hilbert space of \emph{state vectors}. The $\bf_{k,r}$ operators form an orthonormal basis set, and the expectation value of the anticommutator $\langle \{ \mathbf A , \mathbf B \} \rangle$ is the inner product. In the above, we can see that, starting from $\bc_{r}$, we obtain new directions in this operator Hilbert space through the action of the Liouvillian $\mathcal L(\bf) = [\bH, \bf]$, where $\bH$ is the Hamiltonian. 

To obtain the off-diagonal elements of the Green's functions ($r'\neq r$), we use the diagonal Green's functions in an orthonormal polynomial decomposition:
\begin{equation}
G^{(k)}_{r'r}(\omega) = \sum_{k'=0}^{k} m_{k',r'r} \left( Q_{k',r}(\omega) + L_{k',r}(\omega)G^{(k)}_{rr}(\omega) \right).
\label{eq:Green_poly_hybrid}
\end{equation}
The decomposition coefficients $m_{k,r'r}$ are obtained as
\begin{equation}
    m_{k,r'r} = \langle \{\bf_{k,r} , \bc^\dagger_{r'}\} \rangle,
\label{eq:m}
\end{equation}
and $Q_{k,r}(\omega)$ and $ L_{k,r}(\omega)$ are polynomials in $\omega$, both produced by the same three-term recursion relation, 
\begin{equation}
    \beta_{k+1,r}X_{k+1,r}(\omega) = (\omega - \alpha_{k,r}) X_{k,r} - \beta_i X_{k-1,r}
\end{equation}
where $X_{k,r}\in\{L_{k,r},Q_{k,r}\}$ and with initial conditions:
\begin{align}
    L_{-1,r}(\omega) &= 0,\\
    L_{0,r}(\omega) &= 1, \\
    Q_{0,r}(\omega) &= 0, \\
    Q_{1,r}(\omega) &= -1/\beta_{1,r}
\end{align}
From this, we see that the method only requires one additional expectation value~\eqref{eq:m} per iteration, for each off-diagonal element desired.

We use the quantum computer to obtain expectation values \eqref{eq:a}, \eqref{eq:b}, and \eqref{eq:m} 
via repeated statistical sampling of observables in an approximate ground state $|g\rangle = U_g |0\rangle$ prepared with some quantum circuit $U_g$. Such an approach avoids manipulating exponentially large state vectors on classical hardware.

We benchmark the method on the one-dimensional Hubbard model,
\begin{equation}
\begin{split}
    \mathbf{H} =& -t\sum_{i\sigma}\left(\mathbf{c}^\dagger_{i\sigma}\mathbf{c}_{i+1,\sigma} + \mathbf{c}^\dagger_{i+1,\sigma}\mathbf{c}_{i\sigma} \right) 
    - \mu \sum_{i\sigma} \mathbf{c}^\dagger_{i\sigma}\mathbf{c}_{i\sigma} \\
    &+ U \sum_{i} \mathbf{c}^\dagger_{i\uparrow}\mathbf{c}^\dagger_{i\downarrow}\mathbf{c}_{i\downarrow}\mathbf{c}_{i\uparrow},
\end{split}
\label{eq_hubbard}
\end{equation}
where the indices of $\bc_{i\sigma}$ indicate site $i$ and spin $\sigma \in \{\uparrow,\downarrow\}$ explicitly, $t$ is the amplitude of the hopping between sites, $\mu$ is the chemical potential and $U$ is the on-site repulsion.
The energy scale is defined by $t=1$.
We consider the open-boundary four-site case with repulsion $U=4t$ at half filling with ($\mu=U/2$). We convert the Hamiltonian to the Pauli operators representation via the Jordan-Wigner mapping. This results in a Hamiltonian with $17$ terms on $8$ qubits.
Given this small size, we can obtain accurate ground-state circuits at reasonable gate depth, enabling proper benchmarking of the method.

For this benchmarking, we consider three approximate ground states, which we
identify by their fidelity to the true ground state, namely 0.999, 0.963, and 0.768. Their respective
expected energies are $0.999 E_0$, $0.99 E_0$, and $0.9 E_0$, progressively farther from the true ground state energy of $E_0=-9.9531$.
The corresponding preparation circuits are given in Appendix~\ref{ap:A}.
We obtain these circuits by running classical simulations of the variational quantum eigensolver (VQE)~\cite{peruzzo_Variational_2014} algorithm with various ansätze and initial parameters. While VQE is known to suffer from scalability challenges~\cite{larocca_Barren_2025}, 
our algorithm does not depend on this choice for preparing the ground state and is expected to scale to larger systems given an appropriate ground state preparation algorithm. We leave the study of such algorithms to future work.

\begin{figure*}
\includegraphics[width=0.9\textwidth]{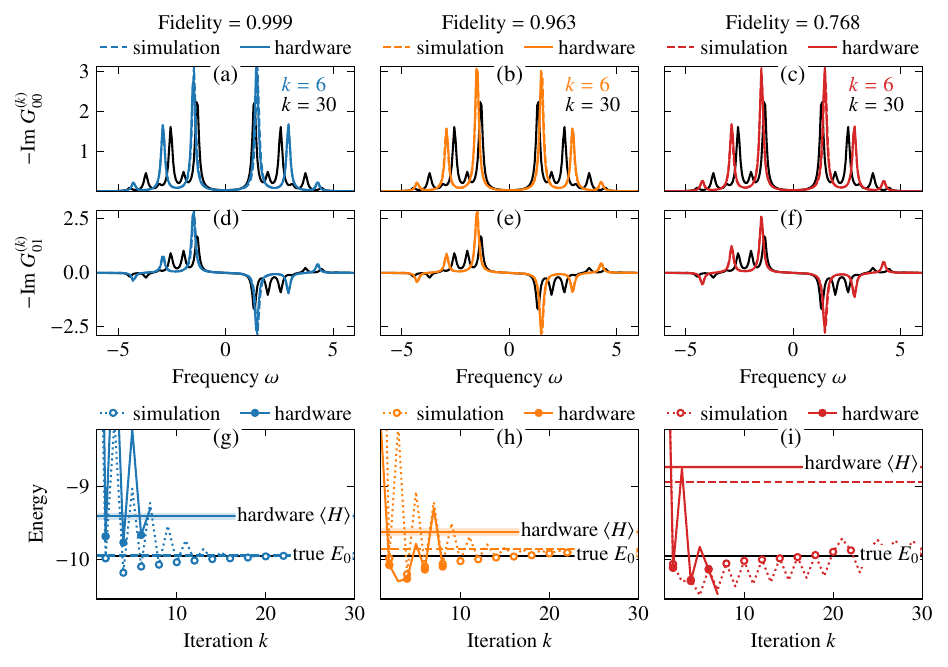}
\caption{
Green's functions and energy estimates obtained with the proposed algorithm. In all panels, black lines are exact values, dashed colored lines are simulation results, and full-colored lines are \emph{IBM Quebec} quantum processor results. In each column, a different approximate ground-state circuit is used for taking expectation values, as identified, at the top, by the circuit's fidelity to the true ground state. 
The first row, (a)-(c), shows the imaginary parts (spectral weight) of the local Green's functions, $G_{00}(\omega)$, obtained with the recursion~\eqref{eq:contfrac}-\eqref{eq:fm1}, while the second row (d)-(f) shows the first-neighbor one, $G_{01}(\omega)$, obtained with polynomial expansion of Eqs.~\eqref{eq:Green_poly_hybrid}. 
Simulation and hardware results are shown for iteration $k=6$, while the exact results (black lines) correspond to iteration $k=30$, at which the algorithm numerically converges. 
The last row, (g)-(i), compares energy estimates from the Galitskii-Migdal formula~\eqref{eq:gm} (jagged colored lines, with markers highlighting even iterations) to the expected values of the Hamiltonian for each circuit (horizontal colored lines, with standard deviation in shaded area) and the true ground state energy (black line). 
}
\label{fig_green}
\end{figure*}

For each circuit, we run classical simulations of our algorithm, as well as hardware demonstrations on IBM's $156$ qubits superconducting processor {\it IBM Quebec}.
We employ the terms \emph{simulation} and \emph{hardware run} to distinguish the two.
Simulations provide exact expectation values for \eqref{eq:a}, \eqref{eq:b}, and \eqref{eq:m}, whereas hardware runs obtain these values through quantum sampling with \emph{Qiskit}'s \emph{Estimator} primitive~\cite{_Qiskit_ibm_runtime_}.
We also worked within the runtime and sample count limitations imposed by the \emph{Qiskit Session} primitive~\cite{_Qiskit_ibm_runtime_}, such that hardware runs do not reach beyond iteration $k=8$.
All hardware runs had precision set to $1/\sqrt{4000}$ (equivalent to 4000 shots) and multiple error mitigation techniques enabled: Twirled Readout Error eXtinction (TREX)~\cite{vandenberg_Modelfree_2022}, Zero Noise Extrapolation (ZNE)~\cite{cai_Multiexponential_2021}, and gate twirling~\cite{wallman_Noise_2016}.
We use automatic qubit selection, and the relevant qubit and gate properties at the time of execution are displayed in Appendix~\ref{ap:A}.

\begin{figure}
\includegraphics{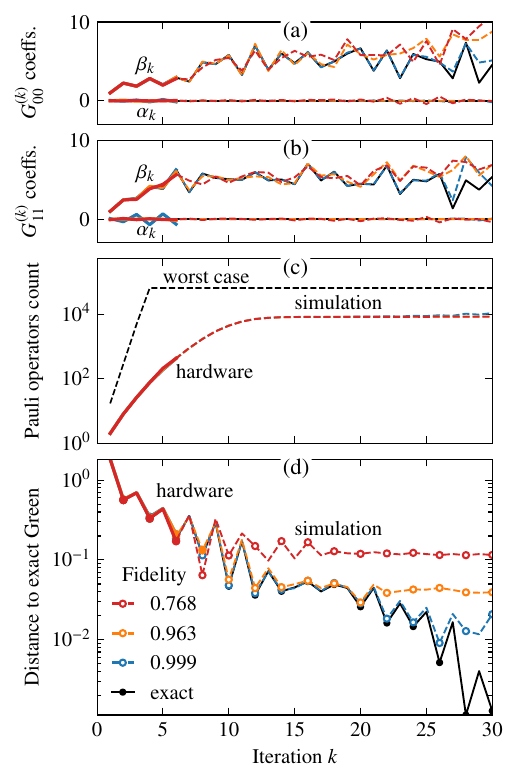}
\caption{
Convergence and computational cost of the proposed algorithm.
In all panels, black lines are exact values, dashed colored lines are simulation results, and full-colored lines are hardware results. The different colors correspond to different approximate ground state circuits.
Panels (a) and (b) show the raw outputs of the algorithm: the continued fraction coefficients obtained at each iteration of the recursion. 
Panel (c) shows the number of Pauli operators generated by the recursion as a function of iterations; it constitutes the main computational cost of the algorithm.
Panel (d) shows the distance~\eqref{eq_wass}, on a logarithmic scale, between the computed Green's function and the exact Green's function, as a function of iterations.
}
\label{fig_wassab}
\end{figure}

Figures~\ref{fig_green}(a)-(f) compare our simulation results (colored dashed line) and our hardware results (full-colored lines) for the Green's functions obtained at iteration $k=6$ of the algorithm. 
The simulation results at k = 30 (black lines) are fully converged and represent the exact Green's function.
Hardware results for the diagonal ($G_{00}$) and off-diagonal ($G_{01}$) Green's functions match their respective simulated values quantitatively. Furthermore, despite using approximate ground state circuits, results at iteration $k=6$ already approach the exact results qualitatively, hinting at the rapid convergence of the algorithm, which we quantify in Fig.~\ref{fig_wassab}(d), as discussed later.

From the Green's functions, we can estimate the ground state energy with the Galitskii-Migdal formula~\cite{galitskii_Application_1958,holm_Total_2000,foley_liouvillian_2024},
\begin{align}
E = \frac{1}{2}\text{Tr}\left[ \oint_{C} \frac{d\omega}{2\pi i} f(\omega)  (\omega + H_0) G(\omega) \right],
\label{eq:gm}
\end{align}
where integration path $C$ encloses all poles of the Green's function, but not the poles of the Fermi-Dirac distribution $f(\omega)$; and $H_0$ is the non-interacting part of the Hamiltonian. For the four-site case considered, 
\begin{align}
    H_0 = - \begin{pmatrix}
        \mu & t & 0 & 0\\
        t& \mu & t & 0 \\
        0 & t& \mu & t\\
        0 & 0 & t&  \mu
    \end{pmatrix}.
\end{align}

Figures~\ref{fig_green}(g)-(i) show the ground state energy estimates obtained from the Green's functions as a function of iterations.
For even iterations~($k=2, 4, 6, ...$), these values are better estimates of the true ground state energy than hardware expected values of the Hamiltonian $\langle H \rangle$ in the approximate ground states.

Even iterations yield better results because the half-filled ($\mu=U/2$) Hubbard model has particle-hole symmetry.
Indeed, the true spectrum is even with respect to $\omega=0$ with no spectral density at $\omega=0$, while a continued fraction with an odd number of coefficients produce an odd number of poles, which must have density at $\omega=0$ when symmetric about $\omega=0$. 
Therefore, the output of the algorithm alternates between gapped and ungapped spectra, which explains the jaggedness from even to odd iterations in all results. 
Here, we show odd iterations for the sake of completeness, but particle-hole symmetry would justify omitting them from the analysis.

With this in mind, we can give a more detailed analysis of Fig.~\ref{fig_green} and how our algorithm improves energy estimation. 
The first approximate ground state circuit, corresponding to Fig.~\ref{fig_green}(g), has a high fidelity of 0.999 such that the simulated $\langle H \rangle$ matches the true energy almost exactly. However, it is a relatively deep circuit of 36 gate layers (see Appendix~\ref{ap:A}) and, due to hardware noise, the hardware $\langle H \rangle$ fails to capture the true ground state value.
The hardware results for the Galitskii-Migdal energies with this circuit also deviate from the simulated ones, but on all even iterations they are closer to the true ground state energy than the hardware~$\langle H \rangle$.
Thus, in this case, the algorithm offers a way to partially circumvent hardware to noise and provide improved energy estimation.
The second circuit, corresponding to Fig.~\ref{fig_green}(h), has a slightly lower fidelity of 0.963 but is shallower, with 15 layers.
Nevertheless, the Galitskii-Migdal energies on iterations 2, 6, and 8 are better than the hardware $\langle H \rangle$. 
Interestingly, the hardware results diverge from simulations at iterations 3 and 5, but return close to simulations at iterations 6 and 8, suggesting inherent noise robustness.
The last circuit, corresponding to Fig.~\ref{fig_green}(i), has a relatively low fidelity of 0.768, such that the simulated~$\langle H \rangle$ and hardware~$\langle H \rangle$ are quite far from the true ground state energy. Despite this, the Galitskii-Migdal energies are all better than~$\langle H \rangle$ and also deviate from simulations at iterations 3 and 5 before reconvening.
These results indicate that the robustness to noise and ground state imperfection studied in Ref.~\cite{irmejs_Approximating_2025} for the Liouvillian recursion carry over to the task of improving energy estimation.

Remarkably, our algorithm produces almost the same results for the higher and lower fidelity ground states.
To understand this, Fig.~\ref{fig_wassab}(a) and~\ref{fig_wassab}(b) present the actual output $\alpha_k$ and $\beta_k$ of the algorithm as a function of iterations. They also compare simulations and hardware runs to an exact run of the algorithm which starts from the true eigenvector.
Hardware results track the simulations closely.
For iterations beyond the reach of hardware, the simulations show that the $\alpha_k$ and $\beta_k$ stemming from lower-fidelity ground states deviate from exact results earlier than the higher-fidelity ones.
Yet, the first few iterations are resilient to ground state quality, or lack thereof. The benefits of an accurate ground state appear after many iterations, which we quantify in Fig.~\ref{fig_wassab}(d), as discussed later.

Figure~\ref{fig_wassab}(c) shows the main computational costs of our approach: the growing number of Pauli operators entering the decomposition of operators $\bf_{k, r\sigma}$. To measure expectation values~\eqref{eq:a},~\eqref{eq:b}, and~\eqref{eq:m}, we must cast operators in the Pauli operator basis,
\begin{align}
    \bf_{k,r} &= \sum_{s} w^{(k, r)}_s \mathbf P_s, & \mathbf P_s \in \{I,X,Y,Z\}^{\otimes n}.
\end{align}
For $n$-qubits, there are $4^n$ possible Pauli operators $\mathbf P_s$. However, we are interested in the number $m$ of those with non-zero weights $w^{(k,r)}_s$. This number determines the quantum computational cost, as the number of circuit executions required to evaluate the expectation values of these $m$ Pauli operators is expected to scale like~$\sqrt{m}$~\cite{reggio_Fast_2024}. It also determines the classical memory and computing requirements, as we need to record the weights $w^{(k,r)}_s$ and compute the commutators and anticommutators in~\eqref{eq:recurs},~\eqref{eq:a},~\eqref{eq:b}, and~\eqref{eq:m} on classical hardware.
Each commutator operation generates new Pauli operators.  Supposing that no repetition nor cancelation occurs, we get a worst-case estimate of ${O}\left(\textrm{min}(h^k,4^n)\right)$ for $k$ iterations, $n$ qubits, and $h$ terms in the Hamiltonian.
Of course, repetitions and cancelations do occur and the actual Pauli operator count as a function of iterations is lower than the worst case, as can be seen in Fig.~\ref{fig_wassab}(c).
Notably, the operator count is the same for all simulations and hardware runs, regardless of the ground state circuit used, indicating that although these cancelations depend on the measured values for $\alpha_k$ and $\beta_k$, hardware noise and ground state imperfections do not change the operators generated significantly.

Despite being lower than the worst-case estimate, the number of operators grows rapidly with iterations. Although Fig.~\ref{fig_wassab}(c) suggests sub-exponential growth, the saturation due to system size is misleading; we investigated the first few iterations for larger systems and confirmed that the growth is exponential (not shown).
In most algorithms, such exponential growth prohibits their usability. However, the recursion method is known to converge exponentially fast~\cite{viswanath_Recursion_1994,irmejs_Approximating_2025}, and in what follows we will show that this fast convergence compensates the exponential cost.

To measure the convergence, we need to track the difference between the computed Green's functions and the exact Green's function as a function of iterations. For reasons given in Appendix~\ref{ap:B}, we opt for the earth-mover distance, also known as the \mbox{Wasserstein} distance. In one dimension, it corresponds to the area between cumulative densities~\cite{vallender_calculation_1974},
\begin{align}
d_k = \int_{-\infty}^{\infty} \bigg| \int_{-\infty}^{\omega}\Im\Big( G^{(k)}_{rr}(\omega') - G^{(\text{exact})}_{rr}(\omega')\Big)\, d\omega' \bigg| d\omega.
\label{eq_wass}
\end{align}
The imaginary parts (spectral weights) are natural candidates to act as a probability densities because they are positive and normalized. Additionally, since the real and imaginary part of the Green's function are related by a Hilbert transform, no information is neglected.

Figure~\ref{fig_wassab}(d) shows how the Wasserstein distance to the true solution decreases exponentially with iteration number.
The distance to solution at which the algorithm ultimately stabilizes depends inversely on the fidelity of the approximate ground state circuit used; the better the ground state, the better the final Green's function. Again, however, all results track one another at low iteration number, and the benefits of having a better ground state circuit only materialize after many iterations.

\begin{figure}
\includegraphics{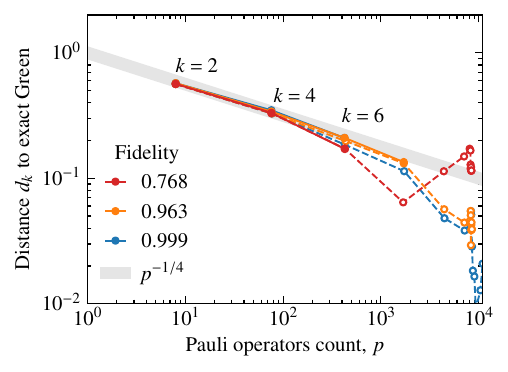}
\caption{
The exponential convergence of the algorithm compensates its exponential cost. The colored lines show the distance $d$ to solution, as measured by~\eqref{eq_wass}, as a function of the number of Pauli operators $p$ generated at even iteration $k$ for simulations (dashed lines) and hardware runs (full lines). The gray line highlights the polynomial scaling $d=p^{-1/4}$.
}
\label{fig_cost}
\end{figure}

Finally, Fig.~\ref{fig_cost} shows that the exponential convergence of the Wasserstein distance with respect to the number of iterations compensates for the corresponding exponential cost in terms of Pauli operators.
The resulting scaling, as shown for even iterations of our hardware runs, is polynomial with a power $1/p^{1/4}$ (pale gray band). Using the Wasserstein distance as a measure of the output error thus suggests a $\mathcal O(1/\epsilon^{4})$ scaling in the computational cost. 
We leave an analytical demonstration of this empirical scaling law to future work.
If this scaling holds for larger systems, and given an efficient approximate ground state preparation method, our algorithm, by measuring expectation values \eqref{eq:a}, \eqref{eq:b}, and \eqref{eq:m} on a quantum computer, would provide exponential advantage to compute Green's functions.

In conclusion, we presented a new quantum algorithm based on the Liouvillian recursion method to compute the electronic Green's functions with a quantum computer.
Results for the four-site Hubbard model on IBM superconducting processors demonstrated rapid convergence of the algorithm with the number of iterations and a significant robustness to noise.
Using the Galitskii-Migdal formula, we were able to obtain better ground state energy estimates than the hardware expectation value of the Hamiltonian, even for a low fidelity approximate ground state.
Although the algorithm requires a number of Pauli operator expectation values exponential in the number of iterations, this cost is compensated for by exponential convergence in the Wasserstein distance, suggesting a computational complexity polynomial in the inverse error.
It remains to be determined if this polynomial scaling and observed noise robustness hold for other systems. 
If this is the case, this work could form the basis of new impurity solvers in dynamical mean field theory (DMFT), which is a clear path to study strongly correlated electrons systems using quantum computers~\cite{selisko_Dynamical_2025,greene-diniz_Quantum_2024}.
Finally, energy estimation using the Galitskii-Migdal formula, as demonstrated here, has many applications beyond the Hubbard model, including spin Hamiltonians and the electronic and vibrational structure of molecules.

\begin{acknowledgments}
Thanks to professors Maxime Charlebois, Stefanos Kourtis, David Sénéchal, André-Marie Tremblay, for insightful discussions.
Thanks to the members of the Algolab Marco Armenta, Tania Belabas, Joshua Cantin, Maxime Dion, Jean-Fréderic Laprade, Ghislain Lefebvre and Christian Sara-Bournet for their help, support and encouragement.
Special thanks to Joshua Cantin and Jean-Fréderic Laprade for suggestions to improve the manuscript.
This work was supported the Ministère de l'économie, de l'innovation et de l'énergie du Québec, the Natural Sciences and Engineering Research Council of Canada (NSERC) under Canada Graduate Scholarships-Master's Program, the Fonds de Recherche du Québec Nature et Technologies under master research scholarships program, and the National Research Council~(NRC) Applied Quantum Computing challenge program.
We acknowledge the use of IBM Quantum services for this work. The views expressed are those of the authors, and do not reflect the official policy or position of IBM or the IBM Quantum team.
\end{acknowledgments}

\bibliography{biblio.bib}

\setlength{\columnsep}{0pt}

\appendix

\setlength{\tabcolsep}{8pt}

\section{Qubits Properties and Circuits}
\label{ap:A}
\begin{figure*}
\subfloat[\emph{Real amplitude} ansatz, 10 repetitions, fidelity$=0.999$, energy$=-9.9487$]{\includegraphics[width=0.96\textwidth]{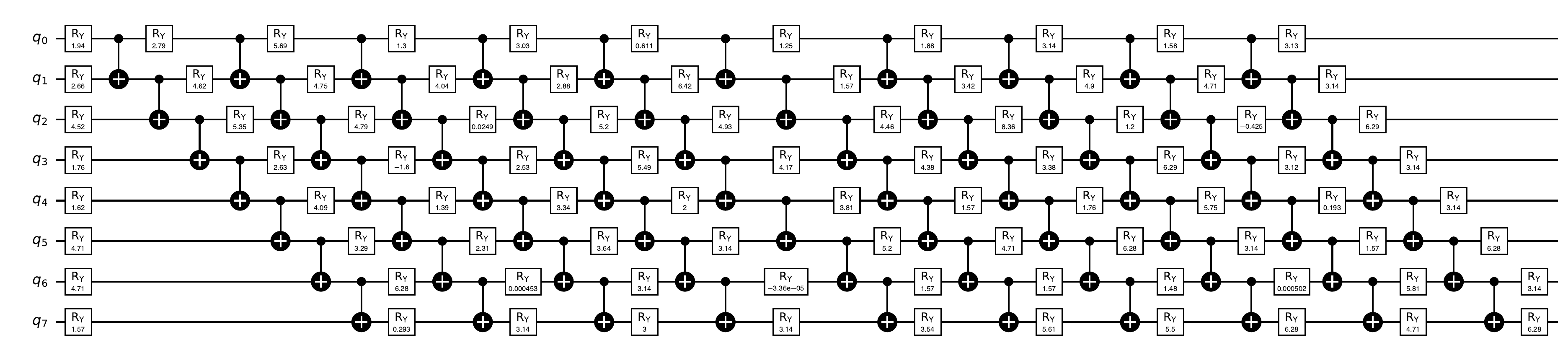}}
\\
\subfloat[\emph{Real amplitude} ansatz, 3 repetitions, fidelity$=0.963$, energy$=-9.8595$]{\includegraphics[width=0.48\textwidth]{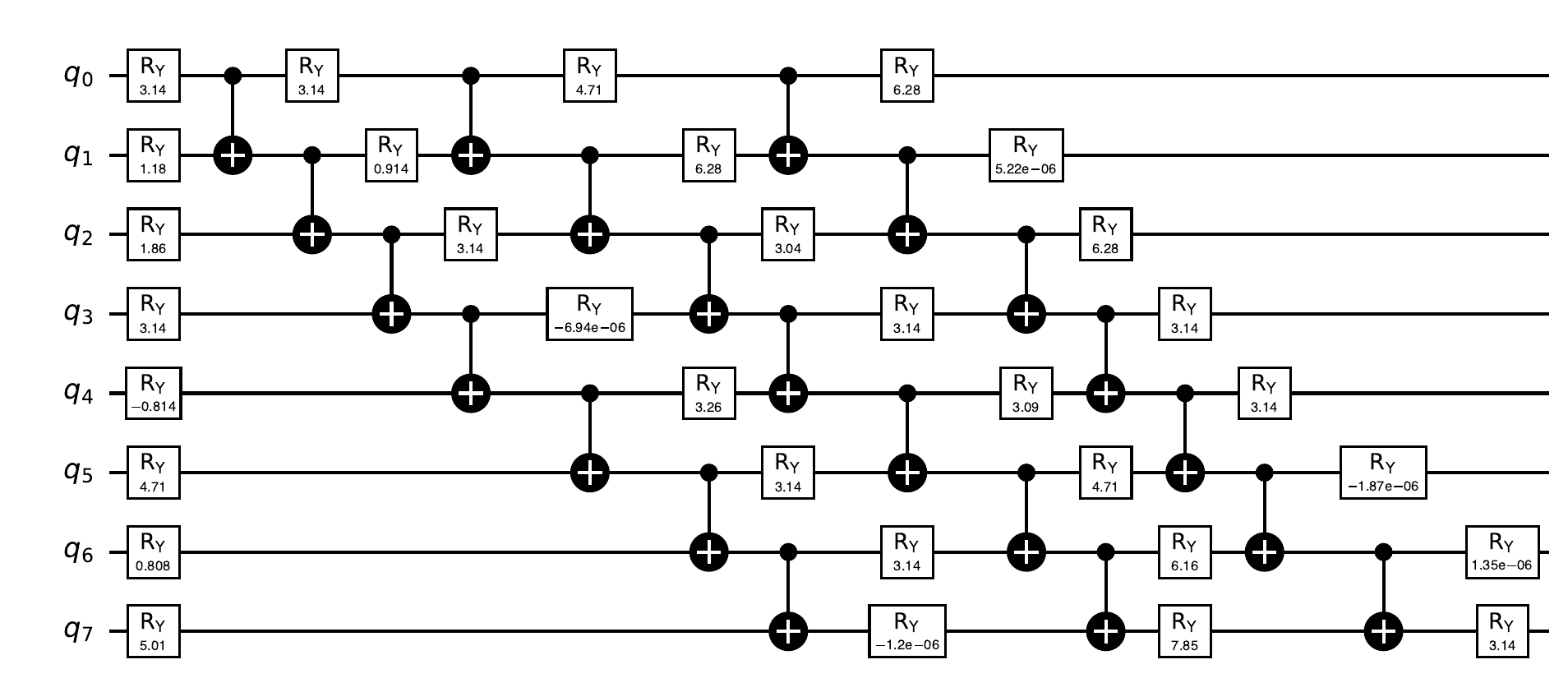}}
\subfloat[\emph{Efficient SU2} ansatz, 3 repetitions, Fidelity=0.768, energy=$-8.9271$]{\includegraphics[width=0.48\textwidth]{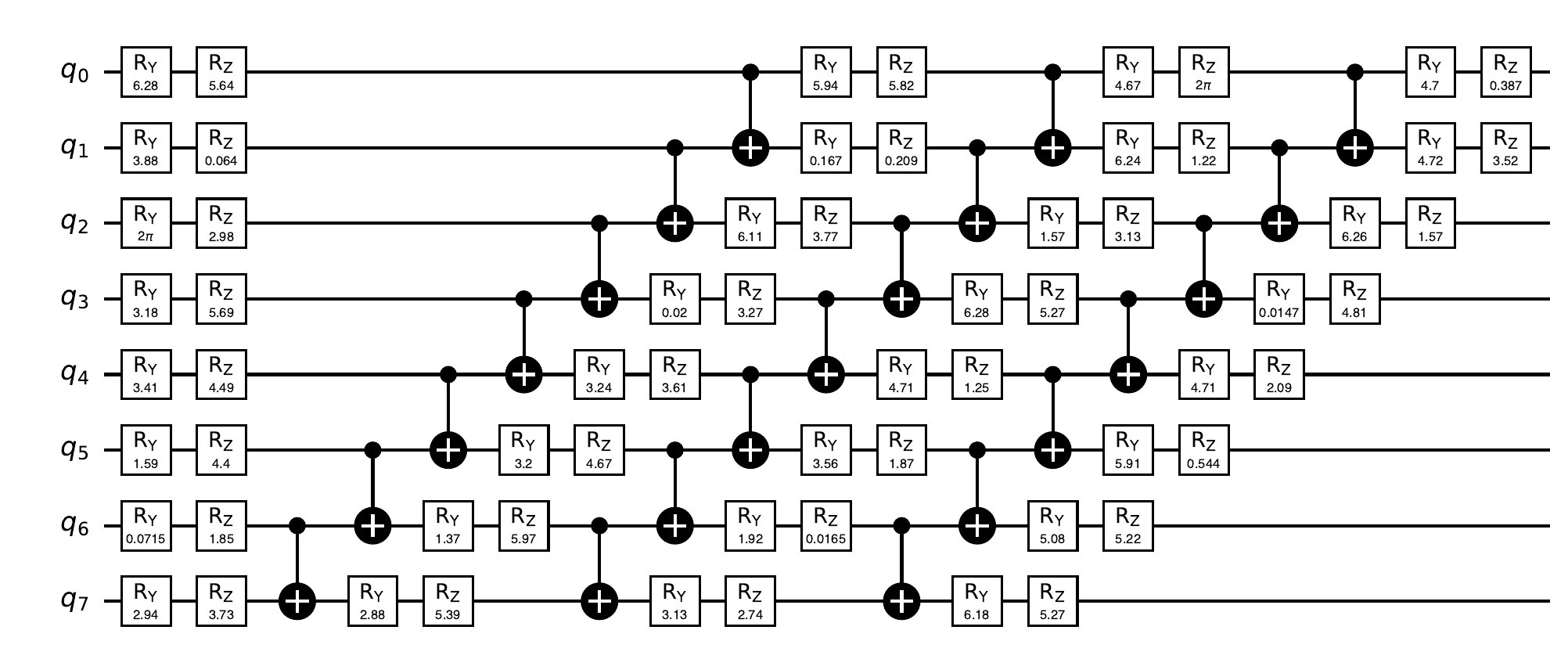}}
\caption{Approximate ground state preparation circuits, progressively farther from the true ground state energy of $-9.9531$.}
\label{fig:circuits}
\end{figure*}

Tables~\ref{table:qubits} and~\ref{table:twogates} present qubits and gates properties during computations.
Figure~\ref{fig:circuits} show the three circuits.

\section{Justification for the Wasserstein Distance}
\label{ap:B}

The Wasserstein distance has one notable advantage compared to other common measures of discrepancy between distributions and functions such as the mean-square error, the Kullback-Leibler divergence, or the total variation distance.
This advantage was highlighted a few years ago by the machine learning community (see reference~\cite{arjovsky_wasserstein_2017}, Fig.~1), and is best described as it providing a ``horizontal'' rather than ``vertical'' measure of the discrepancy.
To clarify what this means, suppose that the two functions consist of a single peak each, without overlap, and separated by some distance $\Delta\omega$. 
Most other measurement strategies are primarily sensitive to the overlap between the two functions, and would yield values that are independent of $\Delta\omega$.
By contrast, the Wasserstein distance is \emph{designed} to be proportional to $\Delta\omega$, and weighted by the amplitude of the peaks.
This sensitivity to distance between peaks is critical to measure the discrepancy of Green's functions.
Furthermore, numerical manipulations usually require an artificial broadening $G(\omega + i\eta)$ with $\eta\rightarrow0$, where $\eta$ acts as a ``resolution'' parameter.
When comparing Green's functions, the effect of increasing~$\eta$ is to artificially increase the overlap. Therefore, most measures of discrepancy such as those mentioned above will change with~$\eta$ meaning that they depend on the resolution used. In contrast, the Wasserstein distance becomes independent of~$\eta$ for small enough~$\eta$, as we verified numerically (not shown), making it a desirable metric for the discrepancy between Green's functions.

\begin{table}[h]
\caption{Qubits properties}
\vspace{5mm}
\label{table:qubits}
\begin{tabular}{cccc}
\toprule
Node & $T_1$ [µs] & $T_2$ [µs] & Readout error \\
\midrule
39 & 265.78 & 323.54 & 0.004639 \\
51 & 277.96 & 254.31 & 0.004150 \\
52 & 323.32 & 340.14 & 0.004517 \\
53 & 255.62 & 223.32 & 0.004395 \\
58 & 184.98 & 243.45 & 0.003052 \\
71 & 256.82 & 294.37 & 0.003418 \\
72 & 177.75 & 213.49 & 0.004150 \\
73 & 252.66 & 276.04 & 0.004761 \\
\bottomrule
\end{tabular}
\end{table}

\begin{table}[h]
\caption{Two-qubit gates properties}
\label{table:twogates}
\vspace{5mm}
\begin{tabular}{ccc}
\toprule
Edge & $CZ$ error & $R_{ZZ}$ error \\
\midrule
(39,53) & 0.001143 & 0.000934 \\
(51,52) & 0.001118 & 0.001488 \\
(51,58) & 0.000792 & 0.001126 \\
(52,53) & 0.001298 & 0.007664 \\
(58,71) & 0.001013 & 0.001425 \\
(71,72) & 0.001152 & 0.001380 \\
(72,73) & 0.000885 & 0.001033 \\
\bottomrule
\end{tabular}
\end{table}

\onecolumngrid

\end{document}